\pdfoutput=1
\documentclass[%
 reprint,
 amsmath,amssymb,
 aps,
 superscriptaddress,
 preprintnumbers,
]{revtex4-2}

\usepackage{graphicx}
\usepackage{dcolumn}
\usepackage{booktabs}
\usepackage{mathtools}
\usepackage{multirow}
\usepackage{mathrsfs}
\usepackage{bm}
\usepackage[caption=false]{subfig}
\newcommand{\overbar}[1]{\mkern1.5mu\overline{\mkern-1.5mu#1\mkern-1.5mu}\mkern 1.5mu}

\def\a{\alpha}

\def\cd{{\cal D}}
\def\ce{{\cal E}}
\def\cf{{\cal F}}

\def\cl{{\cal L}}

\def\cn{{\cal N}}
\def\co{{\cal O}}

\def\car{{\cal R}}

\def\[{\lfloor{\hskip 0.35pt}\!\!\!\lceil}
\def\]{\rfloor{\hskip 0.35pt}\!\!\!\rceil}

\begin{document}

\title{Gauging the R-symmetry of old-minimal $R+R^2$ supergravity}

\author{Yermek Aldabergenov}
\email{yermek.a@chula.ac.th}
\affiliation{Department of Physics, Faculty of Science, Chulalongkorn University, Phayathai Road, Pathumwan, Bangkok 10330, Thailand}
\affiliation{Department of Theoretical and Nuclear Physics, Al-Farabi Kazakh National University, 71 Al-Farabi Ave., Almaty 050040, Kazakhstan}
\affiliation{Institute of Nuclear Physics, 1 Ibraginov Str., Almaty 050032, Kazakhstan}

\date{\today}

\begin{abstract}
Old-minimal $R+R^2$ supergravity has a $U(1)$ R-symmetry which rotates the chiral curvature superfield. We gauge this R-symmetry and study new interactions involving the gauge multiplet in the context of inflation and supersymmetry breaking. We construct models where supersymmetry and the R-symmetry are spontaneously broken during and after Starobinsky inflation, and one-loop gauge anomalies are cancelled by the Green--Schwarz mechanism which can also generate Standard Model gaugino masses. The hierarchy between the auxiliary fields, $\langle F\rangle\gtrsim\langle D\rangle$, leads to split mass spectrum where the chiral multiplet masses are around the inflationary scale ($10^{13}$ GeV), while the gauge multiplet masses can be arbitrarily small.
\end{abstract}

\maketitle

\section{Introduction}

Observations of Cosmic Microwave Background (CMB) fluctuations are in a good agreement with predictions of single-field inflation, and favour models with concave scalar potentials predicting low tensor-to-scalar ratio \cite{Planck:2018jri}. One of these models is the Starobinsky model \cite{Starobinsky:1980te}, which is a theory of $R+R^2$ modified gravity, where the quadratic term in the scalar curvature $R$ gives rise to an additional scalar degree of freedom (called the scalaron) with a particular form of the scalar potential that makes it a good candidate for the inflaton field. 

$N=1$ supersymmetrization of $R+R^2$ gravity is not unique. This is because in standard two-derivative supergravity there are multiple choices of auxiliary fields to complete an off-shell supergravity multiplet. Well-known minimal examples, with $12+12$ degrees of freedom, are old-minimal and new-minimal multiplets \cite{Gates:1983nr}. The former includes a real vector and a complex scalar as auxiliary fields, while the latter includes a real vector and a two-form auxiliary field. In fact, the auxiliary vector of new-minimal supergravity is a gauge field of R-symmetry, $U(1)_R$. This is inconsequential for two-derivative supergravity, because upon eliminating the auxiliary fields, both old-minimal and new-minimal approaches describe the same Einstein supergravity. But if we include higher derivatives, namely an $R^2$ term, the two approaches lead to two different supersymmetric extensions of $R+R^2$ gravity, due to the fact that the auxiliary fields become dynamical. See e.g. Refs. \cite{Kallosh:2013lkr,Farakos:2013cqa,Ketov:2013dfa,Ferrara:2013wka,Ferrara:2014cca} for the realization of inflation in these modified supergravity models.

$R+R^2$ supergravity in the new-minimal formulation can be equivalently described by Einstein supergravity coupled to a massive vector multiplet which gauges the R-symmetry (spontaneously broken everywhere in field space) and includes a real scalar (scalaron) and a massive vector as bosonic degrees of freedom \cite{Cecotti:1987qe}. On the other hand, the old-minimal $R+R^2$ supergravity is equivalent to Einstein supergravity coupled to two chiral (scalar) multiplets \cite{Cecotti:1987sa}. Notably, this theory has global exact R-symmetry, for a suitable choice of K\"ahler potential and superpotential, which rotates one of the chiral scalars (in the higher-derivative formulation, this scalar can be seen as the leading component of the chiral curvature superfield). In this work we gauge the R-symmetry and study the resulting theory in the context of inflation and supersymmetry breaking. 

Supersymmetry breaking in pure old-minimal $R+R^2$ supergravity has been studied for example in \cite{Hindawi:1995qa} where the SUSY-breaking vacuum found by the authors also spontaneously breaks global R-symmetry. This leads to two problems: a massless R-axion, and the fact that the inflationary attractor trajectory generally leads SUSY-preserving vacuum instead of the SUSY-breaking one (see Figure 2 of \cite{Aldabergenov:2020bpt} which shows the scalar potential and inflationary trajectory of this model). In Ref. \cite{Dalianis:2014aya} the authors studied new SUSY-breaking vacua in old-minimal $R+R^2$ supergravity by introducing explicit R-symmetry-breaking terms, which solves both of the above problems. In our approach, these problems are solved by instead gauging the R-symmetry and arranging for its spontaneous breakdown both during and after inflation.

We start in Section \ref{sec_setup} by introducing general old-minimal $R+R^2$ supergravity, and in Section \ref{sec_dual} we describe dual scalar-tensor theories first in terms of the component fields, and then in superspace. We show that one (out of four) real scalar can be integrated out when describing inflation, and obtain convenient form of the effective Lagrangian. In Section \ref{sec_R} we gauge the R-symmetry of the model, and use the resulting theory in Section \ref{sec_inflation} to describe inflation and SUSY breaking in Minkowski vacuum. In Section \ref{sec_anomaly} we study anomaly cancellation condition by the Green--Schwarz mechanism, and obtain the fermion mass spectrum. Finally, in the conclusion section we summarize the results.

\section{Old-minimal $R+R^2$ supergravity}\label{sec_setup}

We start with general ($\cn=1$, $D=4$) old-minimal modified supergravity Lagrangian (throughout the paper we set $M_P=1$ and use the conventions of \cite{Wess:1992cp})
\begin{equation}
    \cl=\int d^2\Theta2\ce\left[\tfrac{1}{8}(\overbar\cd^2-8\car)N(\car,\overbar\car)+\cf(\car)\right]+{\rm h.c.}~,\label{master_L}
\end{equation}
where $\cd^2\equiv\cd^\a\cd_\a$ with supercovariant derivative $\cd_\a$, $\ce$ and $\car$ are density and curvature chiral superfields, respectively. $N(\car,\overbar\car)$ is a real function, while $\cf(\car)$ is holomorphic. Neglecting the fermions, the component expansion of $\ce$ and $\car$ is
\begin{gather}
\begin{gathered}
    2\ce|=e~,~~~\cd^22\ce|=-24e\overbar X~,~~~\car|=X~,\\
    \cd^2\car|=\tfrac{1}{3}R+16X\overbar X-\tfrac{2i}{3}\nabla_mb^m+\tfrac{2}{9}b_mb^m~,
\end{gathered}
\end{gather}
where $|$ extracts $\Theta=0$ component, $e\equiv{\rm det}\,e^a_m=\sqrt{-g}$, $X$ and $b_m$ are complex scalar and real vector auxiliary fields of old-minimal supergravity. The standard Poincar\'e supergravity corresponds to $N=0$ and $\cf=-3\car$, or equivalently $N=-3$ and $\cf=0$. For general function $N(\car,\overbar\car)$, or more specifically if $\partial_\car\partial_{\overbar\car}N\equiv N_{\car\overbar\car}\neq 0$, the theory includes an $R^2$-term ($R$ being the scalar curvature), while $X$ and $b_m$ become dynamical. This can be seen from the component expansion of the Lagrangian \eqref{master_L},
\begin{widetext}
\begin{align}
\begin{aligned}
    e^{-1}\cl &=-\tfrac{1}{12}\left(\cf'+\overbar\cf'+2N+2N_XX+2N_{\overbar X}\overbar X-8N_{X\overbar X}X\overbar X-\tfrac{1}{9}N_{X\overbar X}b^2\right)R\\
    &+N_{X\overbar X}\left[\tfrac{1}{144}R^2-\partial_m X\partial^m\overbar X+\tfrac{1}{36}(\nabla b)^2\right]-\tfrac{i}{3}b_m(N_X\partial^mX-N_{\overbar X}\partial^m\overbar X)\\
    &+\tfrac{i}{6}\nabla b\,(\cf'-\overbar\cf'+2N_XX-2N_{\overbar X}\overbar X)+6X\overbar\cf+6\overbar X\cf+12NX\overbar X\\
    &-\left(\cf'+\overbar\cf'+2N+2N_XX+2N_{\overbar X}\overbar X-4N_{X\overbar X}X\overbar X-\tfrac{1}{18}N_{X\overbar X}b^2\right)\left(4X\overbar X+\tfrac{1}{18}b^2\right)~,
\end{aligned}
\end{align}
\end{widetext}
where $N(\car,\overbar\car)|=N(X,\overbar X)$, $\cf(\car)|=\cf(X)$, and $\cf'\equiv\partial_X\cf$. We also denote $\nabla_mb^m\equiv\nabla b\,$ and $b_mb^m\equiv b^2$, where $\nabla_m$ is the spacetime covariant derivative. It is convenient to introduce the mass scale $M$ of the $R^2$ modification of Einstein supergravity by the redefinitions $X\rightarrow MX/\sqrt{12}$ and $b_m\rightarrow\sqrt{3/2}Mb_m$, and rewrite the Lagrangian as (up to total derivatives)
\begin{align}
\begin{aligned}
    e^{-1}\cl &= \frac{1}{2}\left(A+\frac{1}{3}N_{X\overbar X}b^2\right)R+\frac{N_{X\overbar X}}{12M^2}R^2\\
    &-N_{X\overbar X}\left[\partial_m X\partial^m\overbar X-\frac{1}{2}(\nabla b)^2\right]-Mb_m\Sigma^m\\
    &+\frac{M^2}{2}A\,b^2+\frac{M^2}{12}N_{X\overbar X}b^2b^2-U~,\label{L_R^2}
\end{aligned}
\end{align}
where $A$, $\Sigma_m$, and the Jordan frame scalar potential $U$, are functions of $X,\overbar X$,
\begin{align}
\begin{split}\label{A_def}
    A &=-\tfrac{1}{\sqrt{3}M}(\cf'+\overbar\cf')-\tfrac{1}{3}(N+N_XX+N_{\overbar X}\overbar X)\\
    &\hspace{4.8cm}+\tfrac{4}{3}N_{X\overbar X}X\overbar X~,
\end{split}\\
\begin{split}\label{Sigma_def}
    \Sigma_m &=\tfrac{i}{\sqrt{6}}(N_X\partial_mX-N_{\overbar X}\partial_m\overbar X)\\
    &\hspace{.5cm}+\tfrac{i}{\sqrt{6}}\partial_m\left(\tfrac{\sqrt{3}}{M}\cf'-\tfrac{\sqrt{3}}{M}\overbar\cf'+N_XX-N_{\overbar X}\overbar X\right)~,
\end{split}\\
\begin{split}\label{U_def}
    U &=\tfrac{1}{3}M^2X\overbar X\Big[\tfrac{\sqrt{12}}{M}(\cf'+\overbar\cf')+2(N_XX+N_{\overbar X}\overbar X)\\
    &\hspace{.9cm}-N-4N_{X\overbar X}X\overbar X\Big]-\sqrt{3}M(X\overbar\cf+\overbar X\cf)~.
\end{split}
\end{align}
The bosonic degrees of freedom of this theory are comprised of the complex scalar $X$, one real scalar (the scalaron) from the $R^2$-term, and $b_m$ contributing another real scalar in the form $\nabla b$, which can be seen from its equation of motion (we will derive it in the next section). See also \cite{Hindawi:1995qa,Ketov:2013dfa} for further discussions of the vector $b_m$ in old-minimal $R+R^2$ supergravity.

\section{Dual scalar-tensor theory}\label{sec_dual}

Here we dualize the $R+R^2$ Lagrangian given by \eqref{L_R^2} to scalar-tensor gravity. Although the dualization is often performed in terms of the superfields, since the resulting Lagrangian is that of the standard $N=1$ supergravity coupled to matter, it is nevertheless useful to derive the component dual theory from the Lagrangian \eqref{L_R^2} because it will explicitly separate the Starobinsky-like potential for the scalaron, from the Jordan frame potential $U(X,\overbar X)$ as well as the potential for the effective scalar $\nabla b$. We then find superfield dual theory and compare the results.

\subsection{Component dual}

First, let us write the gravitational part of \eqref{L_R^2} as
\begin{equation}
    e^{-1}\cl_g =\frac{1}{2}\left(A+\frac{1}{3}N_{X\overbar X}b^2\right)R+\frac{N_{X\overbar X}}{12M^2}R^2\equiv\frac{f}{2}~,\label{L_g_R}
\end{equation}
where we have introduced the function $f=f(R,X,\overbar X,b^2)$. We then rewrite $\cl_g$ in terms of the (real) auxiliary field $Z$ as
\begin{equation}
    e^{-1}\cl_g=\tfrac{1}{2}f_Z(R-Z)+\tfrac{1}{2}f~,\label{L_g_f}
\end{equation}
where $f=f(Z,X,\overbar X,b^2)$, and $f_Z\equiv\partial f/\partial Z$. Varying \eqref{L_g_f} w.r.t. $Z$ gives $Z=R$ and leads back to the original Lagrangian \eqref{L_g_R}. On the other hand, via the Weyl rescaling,
\begin{gather}
\begin{gathered}
    g^{mn}\rightarrow f_Zg^{mn}~,~~~e\rightarrow f_Z^{-2}e~,\\
    ef_ZR\rightarrow e\left(R-\tfrac{3}{2}f^{-2}_Z\partial_mf_Z\partial^mf_Z\right)~,
\end{gathered}
\end{gather}
we can bring \eqref{L_g_f} to the Einstein frame where the canonically normalized scalaron $\varphi$ is introduced as $f_Z=\exp(\sqrt{2/3}\varphi)$, and the full bosonic Lagrangian (classically equivalent to \eqref{L_R^2}) reads
\begin{align}
\begin{aligned}
    e^{-1}\cl &= \frac{1}{2}(R-\partial\varphi\partial\varphi)-yN_{X\overbar X}\partial X\partial\overbar X\\
    &+\frac{1}{2}N_{X\overbar X}(\nabla b)^2-My\,b_m\Sigma^m+\frac{M^2}{2}b^2\\
    &-\frac{3M^2}{4N_{X\overbar X}}(1-Ay)^2-y^2U~,\label{L_comp_dual}
\end{aligned}
\end{align}
where we denote $y\equiv\exp(-\sqrt{2/3}\varphi)$.

Let us now look at the equation of motion for $b_m$,
\begin{equation}
    \nabla_m(N_{X\overbar X}\nabla b)+My\Sigma_m-M^2b_m=0~.\label{b_EOM}
\end{equation}
Taking the derivative of \eqref{b_EOM} we obtain
\begin{equation}
    \Box(N_{X\overbar X}\nabla b)+M\nabla_m(y\Sigma^m)-M^2\nabla b=0~,\label{b_EOM_KG}
\end{equation}
which is a Klein--Gordon-like equation for the real scalar field $N_{X\overbar X}\nabla b$ with the mass $M^2/\langle N_{X\overbar X}\rangle$, interacting with $X$ and $\varphi$ through derivative terms. If we identify the scalaron with the inflaton, the mass parameter $M$ is of order Hubble scale. Assuming that the derivative terms are small compared to $M$, i.e. taking the limit $M^2\rightarrow\infty$ in Eq. \eqref{b_EOM} (since it is more restrictive than \eqref{b_EOM_KG}), the first term becomes negligible, and we have
\begin{equation}
    b_m\simeq\tfrac{y}{M}\Sigma_m~,\label{b=Sigma}
\end{equation}
which is reminiscent of algebraic equation of motion for $b_m$ in standard supergravity where it serves as an auxiliary field. Substituting \eqref{b=Sigma} into \eqref{L_comp_dual} and neglecting $(\nabla b)^2\sim M^{-2}$, we obtain the effective Lagrangian
\begin{align}
\begin{aligned}
    e^{-1}\cl_{\rm eff} &= \frac{1}{2}(R-\partial\varphi\partial\varphi)-yN_{X\overbar X}\partial X\partial\overbar X\\
    &-\frac{y^2}{2}\Sigma_m\Sigma^m-\frac{3M^2}{4N_{X\overbar X}}(1-Ay)^2-y^2U~,\label{L_comp_dual_eff}
\end{aligned}
\end{align}
which describes the dynamics of $X$ and $\varphi$. For example the quadratic term $\sim(1-Ay)^2$ is responsible for Starobinsky inflation provided that $X$ is stabilized at $X=0$ by its potential $U$. In general however, $X$ can deviate from zero both during inflation and at the vacuum.

The Lagrangian similar to \eqref{L_comp_dual_eff} was also obtained in \cite{Aldabergenov:2020bpt} for specific choices of $N$ and $\cf$, where the effects of $b_m$ and the R-axion ($\log(X/\overbar X)$) were ignored. The effective Lagrangian \eqref{L_comp_dual_eff} is our new result that holds for general functions $N$ and $\cf$ (with the assumption $N_{X\overbar X}\neq 0$).

\subsection{Superfield dual}

The superfield action \eqref{master_L} can be rewritten with the help of auxiliary chiral superfield $T$ as
\begin{align}\label{L_dual_1}
\begin{aligned}
    \cl=\int d^2\Theta2\ce\big[\tfrac{1}{8}(\overbar\cd^2-8\car)N(S,\overbar S)+\cf(S)\\
    +6T(S-\car)\big]+{\rm h.c.}~,
\end{aligned}
\end{align}
where the original Lagrangian \eqref{master_L} is obtained by varying $T$, which eliminates the chiral superfield $S$ as $S=\car$. To obtain the dual Lagrangian, we use the superfield identity
\begin{align}
\begin{aligned}
    &-6\int d^2\Theta 2{\cal E}{\cal R}T+{\rm h.c.}\\
    &\hspace{1cm}=\tfrac{3}{8}\int d^2\Theta 2{\cal E}(\overbar{\cal D}^2-8{\cal R})(T+\overbar T)+{\rm h.c.}~,
\end{aligned}
\end{align}
and bring \eqref{L_dual_1} to the standard matter-coupled $N=1$ supergravity form,
\begin{equation}
    {\cal L}=\int d^2\Theta 2{\cal E}\left[\tfrac{3}{8}(\overbar{\cal D}^2-8{\cal R})e^{-K/3}+W\right]+{\rm h.c.}~,\label{L_superfield_dual}
\end{equation}
where, after the rescaling $S\rightarrow MS/\sqrt{12}$, K\"ahler potential and superpotential are
\begin{align}
    K &=-3\log\big[T+\overbar T-\tfrac{1}{3}N(S,\overbar S)\big]~,\label{Kahler}\\
    W &=\sqrt{3}MST+\cf(S)~,\label{Superpotential}
\end{align}
such that the complex scalar $S$ (we use the same letter for the superfields $S$ and $T$ and their leading components) is in one-to-one correspondence with $X$ of the $R+R^2$ formulation described by the Lagrangian \eqref{L_R^2}.

The component (bosonic) Lagrangian derived from \eqref{L_superfield_dual} has the familiar form
\begin{align}
    e^{-1}\cl &=\tfrac{1}{2}R-K_{I\bar J}\partial_m\Phi^I\partial^m\overbar\Phi^J-V_F~,\label{L_super_comp}\\
    V_F &=e^K(D_{\bar J}\overbar WK^{\bar JI}D_IW-3W\overbar W)~,\label{V_F_def}
\end{align}
where $K_{I\bar J}=\partial_I\partial_{\bar J}K$ is the K\"ahler metric, $K^{\bar JI}$ is its inverse, and $D_IW\equiv\partial_IW+W\partial_IK$. The indices $I,J$ run through the chiral scalars of the model. In present model we have two such scalars, $\Phi^I=\{T,S\}$. The complex scalar $T$ includes the degrees of freedom associated with the scalaron and the effective scalar from $b_m$ of the Lagrangian \eqref{L_R^2} (or its component-dual \eqref{L_comp_dual}). More precisely, we can introduce the scalaron $\varphi$ through the parametrization
\begin{equation}
    T=\tfrac{1}{2}\Big(t+i\sqrt{\tfrac{2}{3}}\tau\Big)~,~~~t=e^{\sqrt{\frac{2}{3}}\varphi}+\tfrac{1}{3}N(S,\overbar S)~,\label{T_par}
\end{equation}
where the imaginary part $\tau$ describes the same degree of freedom as the effective scalar $\nabla b$ of the $R+R^2$ formulation.

After using \eqref{T_par}, the Lagrangian \eqref{L_super_comp} reads
\begin{align}\label{L_super_comp_2}
\begin{aligned}
    e^{-1}\cl &= \tfrac{1}{2}(R-\partial\varphi\partial\varphi-y^2\partial\tau\partial\tau)-yN_{S\overbar S}\partial S\partial\overbar S\\
    &-\tfrac{i}{\sqrt{6}}y^2(N_S\partial_mS-N_{\overbar S}\partial_m\overbar S)\partial^m\tau\\
    &+\tfrac{1}{12}y^2(N_S\partial_mS-N_{\overbar S}\partial_m\overbar S)^2-V_F~,
\end{aligned}
\end{align}
where, again, $y=e^{-\sqrt{\frac{2}{3}}\varphi}$. Following the same pattern as the previous subsection, we can integrate out $\tau$ taking the limit $M^2\rightarrow\infty$. For this we write down the $\tau$-dependent part of the scalar potential,
\begin{align}\label{V_F_tau}
\begin{aligned}
    V_F(\tau)=M^2y^2\tau N_{S\overbar S}^{-1}\Big[\tfrac{i}{\sqrt{6}}\big(\tfrac{\sqrt{3}}{M}\overbar\cf'-\tfrac{\sqrt{3}}{M}\cf'\\
    +N_{\overbar S}\overbar S-N_SS\big)+\tfrac{1}{2}\tau\Big]~,
\end{aligned}
\end{align}
which leads to its equation of motion,
\begin{align}\label{tau_EOM}
\begin{aligned}
    &\nabla_m(y^2\partial^m\tau)+\tfrac{i}{\sqrt{6}}\nabla_m\big[y^2(N_S\partial^mS-N_{\overbar S}\partial^m\overbar S)\big]\\
    &+M^2y^2N_{S\overbar S}^{-1}\Big[\tfrac{i}{\sqrt{6}}\big(\tfrac{\sqrt{3}}{M}\cf'-\tfrac{\sqrt{3}}{M}\overbar\cf'\\
    &\hspace{3cm}+N_SS-N_{\overbar S}\overbar S\big)-\tau\Big]=0~.
\end{aligned}
\end{align}
When $M^2\rightarrow\infty$, we can integrate out $\tau$ as
\begin{equation}\label{tau_io}
    \tau\simeq\tfrac{i}{\sqrt{6}}\big(\tfrac{\sqrt{3}}{M}\cf'-\tfrac{\sqrt{3}}{M}\overbar\cf'+N_SS-N_{\overbar S}\overbar S\big)~,
\end{equation}
and the resulting effective Lagrangian obtained from \eqref{L_super_comp_2} coincides with the Lagrangian \eqref{L_comp_dual_eff} (after identifying $S$ with $X$).

\section{Gauging the R-symmetry}\label{sec_R}

Having established the effective Lagrangian in the convenient form \eqref{L_comp_dual_eff} (with $\tau$ or $b_m$ integrated out), we now discuss the R-symmetry of the model and its gauging. We can derive the extension of \eqref{L_comp_dual_eff} due to the gauging, by using standard matter-coupled supergravity formulae.

First, let us review the global $U(1)$ R-symmetry of old-minimal $R+R^2$ supergravity in the dual formulation given by Eqs. \eqref{L_superfield_dual}--\eqref{Superpotential}. The main feature of the $U(1)$ R-symmetry, which we call $U(1)_R$, is that it transforms superpotential and the Grassmann coordinate $\Theta$. We use the convention where superpotential and $\Theta$ have the R-charges $q(W)=1$, $q(\Theta)=1/2$,
\begin{equation}
    W\rightarrow We^{i\alpha}~,~~~\Theta\rightarrow\Theta e^{i\alpha/2}~,
\end{equation}
where $\alpha$ is the transformation parameter. By looking at the K\"ahler potential \eqref{Kahler} and superpotential \eqref{Superpotential}, it can be seen that R-symmetry fixes the R-charges of the chiral superfields as $q(T)=0$ and $q(S)=1$ (the curvature superfield $\car$ also has unit R-charge), while the function $\cf$ must be proportional to $S$. We can write it as $\cf=-\sqrt{3}McS/2$, with some constant $c$. Assuming that $c$ is real and positive (the latter is needed for Starobinsky-like inflation), it can be absorbed in Eqs. \eqref{Kahler} and \eqref{Superpotential} by the redefinitions $T\rightarrow cT$, $N\rightarrow cN$, $M\rightarrow\sqrt{c}M$ followed by the constant K\"ahler--Weyl transformation $K\rightarrow K+3\log c$, $W\rightarrow c^{-3/2}W$. As the result, we get the following K\"ahler potential and superpotential without loss of generality,
\begin{align}
    K &=-3\log\big[T+\overbar T-\tfrac{1}{3}N(S\overbar S)\big]~,\label{Kahler_R}\\
    W &=\sqrt{3}MS(T-\tfrac{1}{2})~,\label{Superpotential_R}
\end{align}
where $N(S,\overbar S)=N(S\overbar S)$, as required by R-symmetry.

After gauging the $U(1)_R$, the component Lagrangian reads
\begin{align}
\begin{aligned}
    e^{-1}\cl &=\tfrac{1}{2}R-K_{T\overbar T}\partial_m T\partial^m\overbar T-K_{T\overbar S}\partial_m T\overbar{D^mS}\\
    &-K_{S\overbar T}D_mS\partial^m\overbar T-K_{S\overbar S}D_mS\overbar{D^mS}\\
    &-\tfrac{1}{4}h_RF_{mn}F^{mn}+\tfrac{1}{4}h_IF_{mn}\tilde F^{mn}-V_F-V_D~,\label{L_R_gen}
\end{aligned}
\end{align}
where $F_{mn}=\partial_mA_n-\partial_nA_m$ and $\tilde F^{mn}=\tfrac{1}{2}\epsilon^{mnkl}F_{kl}$ for the $U(1)_R$ gauge field $A_m$, and ordinary derivative $\partial_mS$ has been replaced by the gauge-covariant derivative,
\begin{equation}
    D_mS\equiv \partial_mS-igA_mS~,
\end{equation}
with gauge coupling $g$. Under the $U(1)_R$, $S$ and $A_m$ transform as
\begin{equation}
    S\rightarrow e^{i\alpha(x)}S~,~~~A_m\rightarrow A_m+\tfrac{1}{g}\partial_m\alpha(x)~.
\end{equation}
As for the scalar potential, $V_F$ is given by \eqref{V_F_def} as before, while $V_D$ reads
\begin{equation}
    V_D=\tfrac{1}{2}h_R^{-1}\cd^2~,
\end{equation}
where $\cd$ is the Killing potential of $U(1)_R$,
\begin{equation}
    \cd=gS(K_S+\partial_S\log W)=g(SK_S+1)~.\label{Killing_pot}
\end{equation}
The gauge kinetic function $h$ is generally a holomorphic function of chiral superfields, $h(T,S)$ (we denote $h_R\equiv{\rm Re}\,h$ and $h_I\equiv{\rm Im}\,h$). However, since $T$ appears in \eqref{L_dual_1} as a Lagrange multiplier, in order to keep the modified supergravity structure we take $h$ independent of $T$. On the other hand, tree-level R-symmetry prohibits the $S$-dependence of $h$, but as the model is generally anomalous at one loop (due to R-charged fermions), it is possible to cancel the anomalies by the Green--Schwarz mechanism where $S$-dependent gauge kinetic function is employed, such that the Chern--Simons term (proportional to $h_I$) shifts under $U(1)_R$, cancelling the gauge anomaly, see e.g. \cite{Freedman:2005up,Elvang:2006jk,Antoniadis:2014iea} for more detailed discussions (gravitational anomaly can also be cancelled in a similar fashion). At this stage we take $h=1$, and return to the anomaly cancellation conditions in Section \eqref{sec_anomaly}, where it will be shown that $h=1$ is a good approximation for inflationary models.

Finally, we use the parametrization \eqref{T_par} and integrate out $\tau$ according to \eqref{tau_io}, where the R-symmetry and the choice of $\cf$ leads to $\tau\simeq 0$. Then the Lagrangian \eqref{L_R_gen} becomes
\begin{align}
\begin{aligned}
    e^{-1}\cl_{\rm eff} &=\tfrac{1}{2}(R-\partial\varphi\partial\varphi)-yN_{S\overbar S}D_mS\overbar{D^mS}\\
    &+\tfrac{1}{12}y^2(N_SD_mS-N_{\overbar S}\overbar{D_mS})^2\\
    &-\tfrac{1}{4}F_{mn}F^{mn}-V_{F,{\rm eff}}-V_D~,\label{L_R}
\end{aligned}
\end{align}
where $V_{F,{\rm eff}}$ and $V_D$ are
\begin{align}
\begin{aligned}
    V_{F,{\rm eff}} &=\tfrac{3}{4}M^2N_{S\overbar S}^{-1}(1-Ay)^2+y^2U~,\\
    V_D &=\tfrac{1}{2}g^2(1+yN_SS)^2~.\label{V_eff}
\end{aligned}
\end{align}
$V_D$ is unaffected by integrating out $\tau$, while $V_{F,{\rm eff}}=V_F|_{\tau=0}$. The functions $A=A(S\overbar S)$ and $U=U(S\overbar S)$ are defined in \eqref{A_def} and \eqref{U_def} (taking $X=S$), now with $\cf=-\sqrt{3}MS/2$. The main result of this section is the Lagrangian \eqref{L_R} which we will use to describe inflation and spontaneous SUSY breaking without additional matter fields.

\section{Inflation and SUSY breaking}\label{sec_inflation}

To discuss inflation and SUSY breaking we consider a concrete model where $N$ is of the form
\begin{equation}
    N=S\overbar S-\tfrac{1}{2}\zeta(S\overbar S)^2-\tfrac{4}{9}\gamma(S\overbar S)^3~,\label{N_model}
\end{equation}
where $\zeta$ and $\gamma$ are real constants. This form of $N$ was used in \cite{Aldabergenov:2020bpt} in the context of ultra-slow-roll inflation and primordial black hole production (without SUSY breaking). Since the superpotential \eqref{Superpotential_R} is proportional to $S$, in order to break supersymmetry in Minkowski vacuum we need $\langle S\rangle\neq 0$, which in turn spontaneously breaks $U(1)_R$. Thus, in the broken phase we can use the unitary gauge where the angular part of $S$ (R-axion) is set to zero, and the $U(1)_R$ gauge field becomes massive. We parametrize $|S|=\sigma/\sqrt{2}$, where $\sigma$ is a (almost) canonical real scalar. Then, by using \eqref{N_model} the Lagrangian \eqref{L_R} becomes
\begin{align}
\begin{aligned}
    e^{-1}\cl_{\rm eff} &=\tfrac{1}{2}(R-\partial\varphi\partial\varphi)-\tfrac{y}{2}(1-\zeta\sigma^2-\gamma\sigma^4)\partial\sigma\partial\sigma\\
    &-\tfrac{1}{4}F_{mn}F^{mn}-\tfrac{1}{2}m^2_{A}(\varphi,\sigma)A_mA^m-V~,\label{L_R_final}
\end{aligned}
\end{align}
where the scalar potential is
\begin{align}
\begin{aligned}
    V=V_{F,{\rm eff}} &+V_D=\frac{3M^2(1-Ay)^2}{4(1-\zeta\sigma^2-\gamma\sigma^4)}+y^2U\\
    &+\frac{g^2}{2}\Big[1+\frac{y}{2}\sigma^2\Big(1-\frac{\zeta}{2}\sigma^2-\frac{\gamma}{3}\sigma^4\Big)\Big]^2~.\label{V_final}
\end{aligned}
\end{align}
The functions $A$ and $U$ can now be written as
\begin{align}
A &=1+\tfrac{1}{6}\sigma^2-\tfrac{11}{24}\zeta\sigma^4-\tfrac{29}{54}\gamma\sigma^6~,\\
U &=\tfrac{1}{2}M^2\sigma^2\big(1-\tfrac{1}{6}\sigma^2+\tfrac{3}{8}\zeta\sigma^4+\tfrac{25}{54}\gamma\sigma^6\big)~.
\end{align}
The mass $m_A$ of the vector field is a function of $y$ and $\sigma$,
\begin{equation}
    m^2_A=g^2y\sigma^2(1-\zeta\sigma^2-\gamma\sigma^4)+\tfrac{1}{6}g^2y^2\sigma^4\big(1-\tfrac{\zeta}{2}\sigma^2-\tfrac{\gamma}{3}\sigma^4\big)^2~,
\end{equation}
where the first term comes from the kinetic term $|D_mS|^2$ of \eqref{L_R} as in the usual Abelian Higgs model, and the second term from the second line of \eqref{L_R}. In the $R+R^2$ formulation, the latter term originates from integrating out $b_m$ -- see \eqref{b=Sigma} and \eqref{L_comp_dual_eff}.

\subsection{During slow-roll}

First, let us study the asymptotic form of the potential as $y\rightarrow 0$, or $\varphi\rightarrow\infty$, which corresponds to early inflation,
\begin{equation}
    V=\tfrac{3}{4}M^2(1-\zeta\sigma^2-\gamma\sigma^4)^{-1}+\tfrac{1}{2}g^2+{\cal O}(y)~.
\end{equation}
Here we have two extrema in $\sigma$-direction: $\sigma=0\equiv\sigma_a$ and $\sigma^2=-\zeta/(2\gamma)\equiv\sigma^2_b$. The second derivatives at these points are (also ignoring ${\co}(y)$)
\begin{equation}
    V_{\sigma\sigma}|_a=\tfrac{3}{2}M^2\zeta~,~~~V_{\sigma\sigma}|_b=-3M^2\zeta\Big(1+\tfrac{\zeta^2}{2\gamma}\Big)^{-2}~.\label{Vsigma2}
\end{equation}
For $\zeta\gamma>0$, the only critical point is $\sigma_a$ which is a local minimum (maximum) if $\zeta$ is positive (negative). For $\gamma<0$ and $\zeta>0$, $\sigma_a$ becomes a local minimum and $\sigma_b$ a local maximum (and the potential becomes unbounded from below), and for $\gamma>0$, $\zeta<0$ they switch roles: $\sigma_a$ is a maximum, $\sigma_b$ is a minimum (the potential is well-behaved in this case). The latter choice is suitable for our purposes for the following reason (as opposed to the $\zeta\gamma>0$ case). If $\sigma_a$ (i.e. $\sigma=0$) is a local minimum when $y\rightarrow 0$, the inflationary trajectory will follow the $\sigma_a$ path until $y$ reaches unity, which is always a local minimum of the two-field potential, regardless of the choice of the parameters. At this minimum R-symmetry is unbroken (since $\sigma=S=0$), while SUSY is broken by our D-term cosmological constant $g^2/2$, which is undesirable (SUSY breaking scale cannot be of the same order as the cosmological constant). Therefore we consider the case where $\sigma_a$ is instead a local maximum (at $y\rightarrow 0$) and $\sigma_b$ is a minimum at which $U(1)_R$ is spontaneously broken. This corresponds to $\zeta<0$ and $\gamma>0$. The inflationary trajectory can then follow this $\sigma_b$ path until it reaches a Minkowski minimum at $\sigma\neq 0$ (not necessarily $\sigma_b$) where $U(1)_R$ and SUSY remain broken. This minimum can always be arranged with the suitable choice of the parameters. 

Let us also comment on the stabilization of $\sigma$ during inflation. As can be seen from  \eqref{Vsigma2}, for negative $\zeta$ the effective mass of $\sigma$ (around its local minimum $\sigma_b$) during inflation is proportional to $M\sqrt{|\zeta|}$. Moreover, its kinetic term multiplies a factor of $y$ which is very small at this stage. Therefore $\sigma$ is strongly stabilized during inflation as long as $\zeta$ is not vanishingly small.

As for the inflationary observables $n_s$ and $r$, we can expect the usual prediction of the Starobinsky model,
\begin{equation}
    n_s\simeq 1-2/N_e~,~~~r\simeq 12/N_e^2~,\label{obs_SR}
\end{equation}
with the number of e-folds $N_e$ between $50$ and $60$. This is because the effective scalar potential, after minimizing w.r.t. $\sigma$, can always be written as
\begin{equation}
V=\Lambda-Ze^{-\sqrt{2/3}\hat\varphi}+{\cal O}(e^{-2\sqrt{2/3}\hat\varphi})~,
\end{equation}
where $\Lambda$ and $Z$ are some functions of the parameters $\{\zeta,\gamma,M,g\}$, and $\hat\varphi\equiv\varphi-\langle\varphi\rangle$. Regardless of the values of $\Lambda$ and $Z$, the parameters $n_s$ and $r$ will be given by \eqref{obs_SR} when using slow-roll approximation (assuming that slow-roll is not broken during inflation).

Out of the four parameters $\{\zeta,\gamma,M,g\}$, the mass parameter $M$ is fixed by the CMB value of the amplitude of scalar perturbations, $A_s=2.1\times 10^{-9}$ \cite{Planck:2018jri}, and one other parameter, say $\gamma$, is fixed by Minkowski vacuum equations $V=V_\varphi=V_\sigma=0$ (which we solve numerically). Hence, we have two free parameters $\zeta$ and $g$, but with restricted domains. We choose $\zeta<0$ (and $\gamma>0$) as mentioned earlier, and $g/M\leq\co(1)$ because too large $g$ can spoil (F-term-driven) inflation.

\begin{figure}
\centering
  \centering
  \includegraphics[width=1\linewidth]{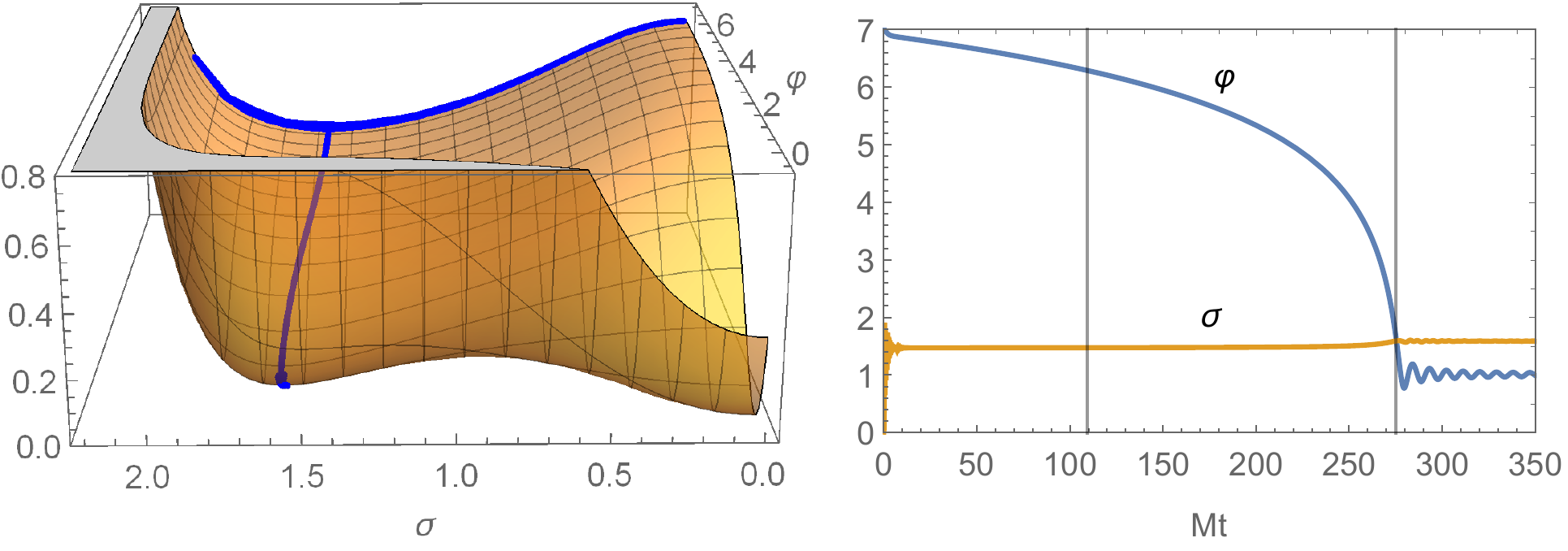}
\captionsetup{width=1\linewidth}
\caption{Left: trajectory of the inflationary solution superimposed on the scalar potential $V/M^2$. Right: evolution of the scalars $\varphi(t)$ and $\sigma(t)$ with normalized time $Mt$. Initial conditions are $\varphi(0)=7$, $\sigma(0)=\dot\varphi(0)=\dot\sigma(0)=0.01$; the vertical lines show the start and end of the last $55$ e-folds.}\label{Fig}
\end{figure}

Let us demonstrate the scalar potential and inflationary solution by fixing the parameters,
\begin{equation}
    \zeta=-1~,~~~\gamma=0.232~,~~~g/M=0.1~,
\end{equation}
where $M=1.77\times 10^{-5}$. We then numerically solve equations of motion for $\varphi(t)$ and $\sigma(t)$ in FLRW spacetime $g_{mn}={\rm diag}(-1,a^2,a^2,a^2)$, where $a$ is time-dependent scale factor. The inflationary solution is shown in Figure \ref{Fig}, where we take the initial conditions as $\varphi(0)=7$, $\sigma(0)=\dot\varphi(0)=\dot\sigma(0)=0.01$. It can be seen that given a small perturbation of $\sigma$ around zero, it will quickly fall into its local minimum where scalaron-driven slow-roll inflation begins. Assuming the observable inflation lasts $55$ e-folds, we calculate the values of the spectral tilt $n_s$ and tensor-to-scalar ratio $r$ at the horizon exit,
\begin{equation}
    n_s=0.9650~,~~~r=0.0036~,
\end{equation}
which are in agreement with CMB data, and indistinguishable from the predictions of single-field Starobinsky inflation.

After inflation, the fields start oscillating around the Minkowski vacuum at $\langle\varphi\rangle\approx 1.01$ and $\langle\sigma\rangle\approx 1.59$. Figure \ref{Fig} (left) also shows the additional local minimum at $\varphi=\sigma=0$, which is de Sitter since at this point we have $V=g^2/2$. Therefore our Minkowski minimum at $\sigma\neq 0$ is stable. In this example, the masses of $\varphi$ and $\sigma$ around the SUSY breaking minimum are $m_\varphi\approx 0.7M$ and $m_\sigma\approx 1.44M$.

\subsection{Spontaneous breaking of SUSY and $U(1)_R$}

Supersymmetry breaking scale is characterized by the auxilary field values at the minimum, as well as the gravitino mass $\langle m_{3/2}\rangle=\langle e^{K/2}|W|\rangle$. Equations of motion for the auxiliary $F$-fields yield
\begin{equation}
    F^I=-e^{K/2}D_{\bar J}\overbar WK^{\bar JI}~.
\end{equation}
In the standard SUGRA formulation of our model where $K$ and $W$ are given by \eqref{Kahler_R} and \eqref{Superpotential_R}, we have two auxiliary $F$-fields (using the parametrization \eqref{T_par} for $T$),
\begin{align}
    F^T &=\frac{M}{2}\sqrt{\frac{y}{3}}\bigg[(N-3)\Big(\overbar S-\frac{N_S}{3N_{S\overbar S}}\Big)\nonumber\\
    &\hspace{1cm}-\frac{2N_SN_{\overbar S}\overbar S}{3N_{S\overbar S}}+\frac{1}{y}\Big(\overbar S-\frac{N_S}{N_{S\overbar S}}\Big)\bigg]~,\\
    F^S &=\frac{M}{2N_{S\overbar S}}\sqrt{\frac{y}{3}}\Big(3-N-N_SS-\frac{3}{y}\Big)~,
\end{align}
and one $D$-field from the $U(1)_R$ gauge multiplet,
\begin{equation}
    D=-\cd=-g(1+yN_SS)~,
\end{equation}
where $\cd$ is the Killing potential \eqref{Killing_pot}. Since we assume $g\lesssim M$, SUSY breaking is dominated by the F-terms which are both non-zero, and are of order Hubble scale.

Let us consider three examples with $\zeta=\{-0.1,-1,-10\}$ keeping $g=0.1M$, and compute the auxiliary field VEVs. The results are presented in Table \ref{Tab} where we also include the values of $\gamma$ found from the vacuum equations for each choice of $\zeta$. It can be seen that the values of the F-terms becomes smaller as we increase $|\zeta|$, while the D-term becomes only slightly smaller. In particular for $\zeta=-10$ the F-terms and the D-term are of the same order if $g=0.1M$. Since $g$ has no lower bound, we can take much smaller values such that the F-terms always dominate.

\begin{table}[hbt!]
\centering
\begin{tabular}{c | r r r}
$\zeta$~ &~ $-0.1$ &~ $-1$ &~ $-10$ \\
\hline
$\gamma$~ &~ $0.013$ &~ $0.232$ &~ $5.898$ \\
$|\langle F^T\rangle|$ &~ $0.7M$ &~ $0.39M$ &~ $0.21M$ \\
$|\langle F^S\rangle|$~ &~ $2.26M$ &~ $0.95M$ &~ $0.33M$ \\
$|\langle D\rangle|$~ &~ $2.56g$ &~ $1.98g$ &~ $1.76g$ \\
$\langle m_{3/2}\rangle$~ &~ $1.03M$ &~ $0.54M$ &~ $0.29M$ \\
\end{tabular}
\captionsetup{width=1\linewidth}
\caption{VEVs of the auxiliary fields and the gravitino mass, in Minkowski vacuum. $\zeta$ are chosen by hand, while $\gamma$ are found from the vacuum equations (for $g=0.1M$).}
\label{Tab}
\end{table}

To complete the picture let us also write down the masses of $\varphi$, $\sigma$, and $A_m$:
\begin{align}
\begin{aligned}
    m_\varphi/M &\approx 0.94~,~0.70~,~0.45~,\\
    m_\sigma/M &\approx 1.39~,~1.44~,~1.44~,\\
    m_A/g &\approx 2.18~,~1.71~,~1.54~,\label{m_scalar}
\end{aligned}
\end{align}
for $\zeta=-0.1$, $-1$, and $-10$, respectively. Although we used $g=0.1M$ to obtain the vacuum values of the scalar fields, we keep the gauge coupling $g$ in Eq. \eqref{m_scalar} and Table \ref{Tab}, to show explicitly the proportionality of $m_A$ and $\langle D\rangle$ to $g$.

\section{Anomaly cancellation and fermion masses}\label{sec_anomaly}

Both chiral fermions of the model, which we call $\chi^T$ and $\chi^S$, as well as the R-gaugino $\lambda$ and the gravitino $\psi$, carry non-zero R-charges,
\begin{equation}
    q(\chi^T)=-1/2~,~~~q(\chi^S)=q(\lambda)=q(\psi)=1/2~,\label{q_fermi}
\end{equation}
which leads to gauge and gravitational anomalies at one loop. These anomalies can be cancelled by the Green--Schwarz mechanism, where a set of appropriate Chern--Simons terms is added to the Lagrangian, such that their gauge transformations cancel the anomalies \cite{Freedman:2005up,Elvang:2006jk,Antoniadis:2014iea}. In particular, for the cancellation of the $[U(1)_R]^3$ anomaly we employ the $S$-dependent gauge kinetic function \cite{Antoniadis:2014iea},
\begin{equation}
    h=1+\beta\log S~,~~~\beta\equiv -\frac{g^2C_R}{12\pi^2}~,\label{h_anomaly}
\end{equation}
where $C_R$ is determined by the R-charges $q$ of the fermions,
\begin{equation}
    C_R={\rm Tr}(q^3)=q(\chi^T)^3+q(\chi^S)^3+q(\lambda)^3+3q(\psi)^3~.
\end{equation}
Using \eqref{q_fermi} we obtain $C_R=1/2$, and the resulting Chern-Simons term has the necessary transformation property under the $U(1)_R$,
\begin{equation}
    \tfrac{1}{4}h_IF_{mn}\tilde F^{mn}\rightarrow \tfrac{1}{4}(h_I+\alpha\beta) F_{mn}\tilde F^{mn},
\end{equation}
since $S$ transforms as $S\rightarrow e^{i\alpha}S$. Notice that the log-term of $h$ in \eqref{h_anomaly} is proportional to $g^2$, which makes it negligible because $g\lesssim M=\co(10^{-5})$ (using Planck units) and $|S|=\sigma/\sqrt{2}$ generally stays at $\co(1)$ during and after inflation. Therefore we can safely use $h\approx 1$ when studying inflationary dynamics.

Once Supersymmetric Standard Model (SSM) is added to the picture, it will bring additional R-charged fermions, depending on how the SSM superfields are charged. For example, a fermion of any neutral chiral superfield has the R-charge of $-1/2$, while a gaugino has R-charge $1/2$. This leads to mixed anomalies $G_{\rm SSM}\times U(1)_R$ which can be cancelled, similarly to the $[U(1)_R]^3$ anomaly, by implementing $S$-dependent gauge kinetic matrix of the Standard Model. The diagonal elements of the gauge kinetic matrix will include terms $\sim g_a^2\log S$ (for Standard Model gauge couplings $g_a$), which would generate SSM gaugino masses, often one or two orders of magnitude smaller than the gravitino mass \cite{Antoniadis:2014iea,Aldabergenov:2021uye}. We leave the implementation of the SSM in our model for future work, and below we consider only the fermions $\chi^T$, $\chi^S$, and $\lambda$.

Since all three multiplets contribute to SUSY breaking, the goldstino $\eta$ is a linear combination,
\begin{equation}
    \eta=\partial_I(K+\log W)\chi^I-\frac{ig\cd}{\sqrt{2}W}e^{-K/2}\lambda~.
\end{equation}
In the unitary gauge $\eta=0$ we are left with two physical massive spin-$1/2$ fermions. After fixing the parameters $\zeta$ and $\gamma$ as in Table \ref{Tab} with $g=0.1M$, and diagonalizing the kinetic and mass matrices, we obtain the following masses for the two Weyl fermions (at the Minkowski vacuum),
\begin{align}\label{fermi_masses}
\begin{aligned}
    m_1/M &\approx 1.00~,~0.72~,~0.45~,\\
    m_2/M &\approx 0.07~,~0.08~,~0.11~,
\end{aligned}
\end{align}
for $\zeta=-0.1$, $-1$, and $-10$, respectively. For larger $|\zeta|$ we can see that the gap between the two masses becomes smaller: $m_1$ decreases while $m_2$ slightly increases.

As we decrease $g$, the heavier fermion mass $m_1$ is unchanged and given by \eqref{fermi_masses}. On the other hand, $m_2$ is proportional to $g^2/M$, which is consistent with the limit $g\rightarrow 0$ where the R-gaugino becomes massless, because the mass term of $\lambda$ is proportional to $F^S\partial_Sh$ (this contains $g^2$), while the mixing terms $\lambda\chi$ are proportional to $g$. As there is no lower limit on $g$, in principle the lighter physical fermion (which is dominated by $\lambda$ for small $g$) can be arbitrarily light.

\section{Conclusion}

In this work we studied a new class of old-minimal $R+R^2$ supergravity models with gauged R-symmetry in the context of inflation and supersymmetry breaking. We started from general (ungauged) old-minimal $R+R^2$ supergravity which is equivalent to Einstein supergravity coupled to two chiral multiplets. For convenience we derived a simplified effective Lagrangian \eqref{L_comp_dual_eff} by integrating out an irrelevant heavy scalar (sinflaton) $\tau$, or in the higher-derivative formulation $\nabla b$. We then gauged the R-symmetry by introducing an abelian vector multiplet, and studied inflation and SUSY breaking vacua in a simple example where K\"ahler potential is given by \eqref{Kahler_R} and \eqref{N_model}. The model has one mass parameter $M$ from the superpotential \eqref{Superpotential_R}, which is fixed by the inflationary scale, the $U(1)_R$ gauge coupling $g$, and parameters from the K\"ahler potential, of which there are two in our example.

Inflation is effectively single-field Starobinsky-type, driven mainly by the F-term, and consistent with CMB data, while SUSY can be broken in Minkowski vacuum by both F- and D-terms. R-symmetry can be spontaneously broken before the onset of observable inflation, and remain broken at the Minkowski vacuum. This leads to the Higgs mechanism where the gauge field becomes massive by combining with the R-axion. Of the two remaining dynamical scalars, one ($\sigma$) is responsible for the aforementioned R-symmetry breaking, while the other one -- the scalaron $\varphi$ -- drives inflation. Because large D-term can spoil inflation, it must be bounded from above, $\langle D\rangle\lesssim \langle F\rangle$, or in terms of the parameters, $g\lesssim M$.  This in turn creates split mass spectrum after SUSY and $U(1)_R$ breaking: three real scalars $\{\varphi,\sigma,\tau\}$, one physical spin-$1/2$ fermion, and the gravitino have masses of order $M$, and on the other side the vector boson and the second spin-$1/2$ fermion have masses of order $g$ and $g^2/M$, respectively.

Cubic anomalies due to the non-zero R-charges of the fermions can be cancelled by the Green--Schwarz mechanism, which requires field-dependent gauge kinetic function transforming under the $U(1)_R$. Once the visible sector is added, the same mechanism will also introduce field-dependence to the Standard Model gauge kinetic matrix, which will give masses to the gauginos. In future works it would be interesting to study reheating, addition of Supersymmetric Standard Model, and dark matter candidates in this setup.

\begin{acknowledgments}
This work was supported by by CUniverse research promotion project of Chulalongkorn
University (grant CUAASC), Thailand Science research and Innovation Fund Chulalongkorn University CU$\_$FRB65$\_$ind (2)$\_$107$\_$23$\_$37, and by the Science Committee of the Ministry of Education and Science of the Republic of Kazakhstan (Grant \# BR10965191 ``Complex research in nuclear and radiation physics, high-energy physics and cosmology for development of the competitive technologies").
\end{acknowledgments}

\end{document}